# Femtosecond laser processing for blast-hole analysis: laser removal of slurry and effect on rocks

Julia Brand[1], Ksenia Maximova[1], Steve Madden,[1] Andrei V. Rode[1], Ehsan Mihankhah[2], John Zigman[2], Andrew J. Hill[2] and Ludovic Rapp[1*]

[1]Laser Physics Centre, Department of Quantum Science and Technology, Research School of Physics, Australian National University, Canberra, ACT 2600, Australia
[2] Australian Centre for Robotics, Faculty of Engineering, University of Sydney, NSW 2006, Australia
[*]ludovic.rapp@anu.edu.au

**Abstract:** This study investigates the possibility of using a femtosecond pulse laser to remove iron ore slurry used to stabilise blast-hole structures by mining industries, intending to preserve the wall's stability and the chemical and compositional properties of the underlying rock. In situ minerals are often coated in other material deposits, such as dust or slurry in blast holes. To analyse the rock materials beneath, its surface must be exposed by removal of the surface layer. The ablation depth per pulse and ablation efficiency of the slurry were determined using femtosecond laser pulses. Then, the ablation of rocks of economic interest in Australia, including banded iron, limonite, goethite, shale, and hematite, was studied to establish their ablation thresholds and rates. Any damage induced by the laser was investigated by optical microscopy, optical profilometry, colourimetry, VIS/NIR spectroscopy and Fourier Transform Infrared spectroscopy (FTIR).

1. ## Introduction

The main iron ores in Australia, contributing to its position as one of the world's leading iron ore producers are hematite and magnetite [1]. Hematite (of chemical formula $Fe_2O_3$) contains about 70% of iron by molecular weight and has been the dominant iron ore mined in Australia since the early 1960s. Approximately 96% of Australia's iron ore exports are high-grade hematite, the majority mined from deposits in the Hamersley province of Western Australia. Most of the world's iron ores occur in banded iron formations (BIF), a sedimentary rock with alternative layers of iron-rich minerals (usually magnetite or hematite) and silica (chert) [2]. These formations are the most important concerning resources and production, and the iron content varies widely. Pisolitic limonite deposits are the next in importance because of their low number of impurities. They are not as rich in iron as the BIF ores. Those mined usually contain 57%-59% iron [2]. Goethite (FeO(OH)) is a common ore in Australia, especially in weathered and lateritic deposits. It often occurs alongside hematite but contains a lower iron content. It can be found in Western Australia, typically mixed with hematite or magnetite.

Drill and blast are commonly used to break up hard rocks in open-pit and underground iron ore mining operations. Although studying the ores is not the primary purpose of these operations, knowing the ore properties in the blast holes facilitates subsequent mine planning, grade control and management of safe operations. Determining the geological properties of the rocks at different locations and depths *in situ* can provide essential insights into the areas of interest, informing: accurate boundaries for grades, short-term planning, and geotechnical purposes [3-8].

The drill-and-blast method involves drilling into a rock, filling it with explosives, and controlled blasting to facilitate rock excavation. Blasthole drilling typically results in a mixture of crushed rocks and fine particles from the hole. Most of this mixture is pushed out of the blast hole and forms a cone above the hole. Some of this mixture, combined with water, covers the wall of the blast hole like a thick layer of slurry, and some of this slurry overflows to cover the surface of the blast hole cones too. The coating (which is referred to as slurry in the rest of this document) provides structural support for friable or waterlogged materials when dried. The thickness of the slurry can vary from a few mm to several cm, and the particles of rocks, clays and ores from the holes comprise an uncontrolled distribution up to approximately 5 cm diameter, though typically in the order of 2 cm. The presence of the slurry prevents direct examination of the rock surface and depth profiling down the holes, and its make-up is a mixture from the entire hole and not necessarily representative of the rock behind it (See Fig. 1).

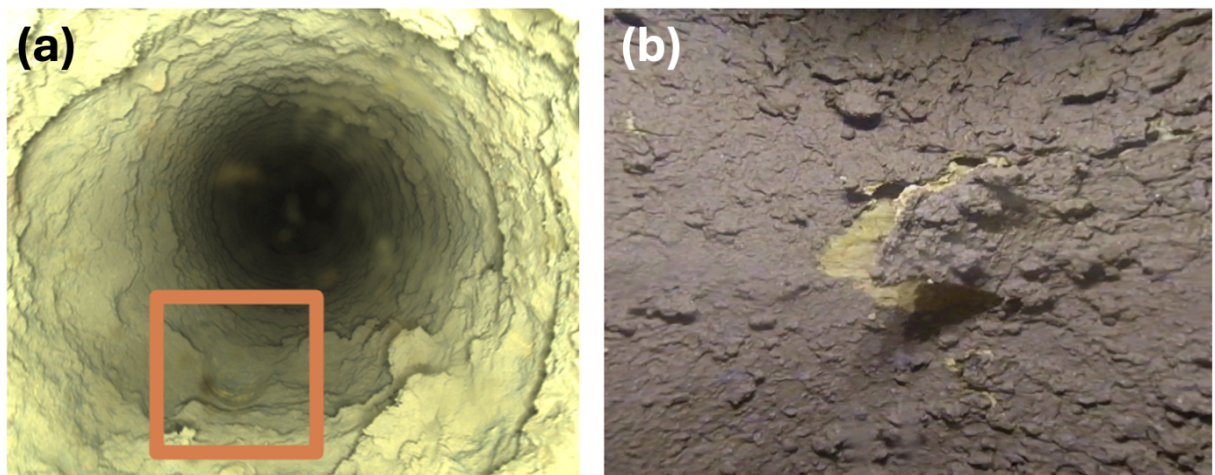

Fig. 1. (a) Blasthole drilling produces a mixture of crushed rocks, fine particles, and water that covers the wall of the holes like a thick layer of slurry. The thickness of the slurry can vary from a few mm to several cm. The box highlights a few cm of the slurry that was scrapped off the wall. (b) The presence of the slurry prevents direct examination of the rock surface.

Therefore, finding a contactless method to remove the slurry from the blast hole to provide access to the rock without disturbing the wall's stability or changing the chemical and compositional properties of the underlying rock is of significant interest.
Recently, femtosecond pulse lasers have emerged as a potential solution for ablating mineralogical materials. Femtosecond lasers are exceptionally effective for ablating stones due to their ability to deliver ultrashort, high-intensity pulses that precisely target and remove materials without causing significant thermal damage to surrounding areas [9, 10]. This "cold" ablation process minimises heat transfer [11], preventing the minerals from melting or adversely affecting nearby structures. Femtosecond pulses offer enough precision for selective material removal, making removing unwanted layers without damaging the underlying substrate possible [12]. The laser ablation of contaminants from rocks using femtosecond pulses has already been demonstrated, for example, the removal of paints and graffiti from granite [13-16], dirt from marble [17,18], or biological colonisations [19-23] and crusts [24-26]. Although the technique has proven effective, the interaction between the laser and the rock can generate side effects depending on the laser wavelength used and the laser intensity, such as discolouration and melting of some minerals. These effects are particularly remarkable in micas (biotite, muscovite), and sometimes feldspars [27]. The sensitivity of biotite has been reported on many occasions, and shortening the pulse duration of the laser was identified as a potential solution to mitigate the damage [12, 28].

This study explores the use of a femtosecond pulse laser emitting in the infrared wavelength domain (IR, 1030 nm) to remove slurry patches while preserving the integrity of underlying rocks for subsequent analysis. The rocks selected for this study included banded iron, limonite, goethite, shale, and hematite. As explained above, banded iron, pisolitic limonite, goethite and hematite are economically significant in Australia in the iron ore mining industry. Shale often accompanies iron ore deposits and can act as a host rock. It is usually a waste material of mining. The material that composes the studied slurry mix was taken from the overflow of the slurry that covers the surface of the blast hole cones of a BIF (Banded Iron Formation) deposit from the Brockman Iron Formation in Western Australia. That ensures the studied sample is highly representative of real-world composition.

First, we determined the laser ablation parameters for the ablation of slurry and rocks, including the identification of ablation thresholds, rock alteration thresholds, ablation rates, and efficiencies. Visual assessment was conducted using optical microscopy, optical profilometry, and colourimetry. Then, the effects of exposure of the rock samples to the laser were examined using visible/near-infrared (VIS/NIR) spectroscopy and Fourier Transform Infrared spectroscopy (FTIR). Finally, we determined the most efficient ablation conditions to remove slurry from goethite using femtosecond pulses, achieving high-rate material removal with minimal alteration of the underlying rock and minimal disruption of the slurry surrounding the ablation window.

## 2. Materials and methods

### a. Approach

The study aims at removing slurry from blastholes to provide access to the rock without disturbing the hole wall's stability or changing the chemical and compositional properties of the underlying rock. As mentioned above, the laser parameters for the ablation of slurry and rocks were first determined, including the identification of ablation thresholds, rock alteration thresholds, ablation rates, and efficiencies. Visual assessment was conducted using optical microscopy, optical profilometry, and colourimetry. Then, the effects of exposure of the rock samples to the laser were examined using visible/near-infrared (VIS/NIR) spectroscopy and Fourier Transform Infrared spectroscopy (FTIR). Finally, we determined the most efficient ablation conditions to remove slurry from goethite using femtosecond pulses, achieving high-rate material removal with minimal alteration of the underlying rock and minimal disruption of the slurry surrounding the ablation window, facilitating accurate analysis, characterisation and identification of the rock. The diagram in Fig. 2 illustrates the process of using a femtosecond pulse laser to open a section of slurry and reveal the rock underneath.

## 1- Preparation
Application of slurry on stone

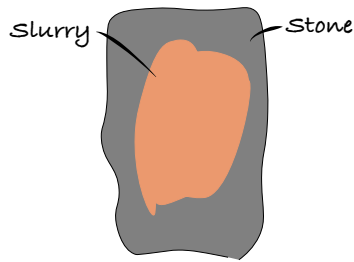

## 2- Laser removal
Opening a patch in the slurry using femtosecond laser pulses to expose the stone underneath

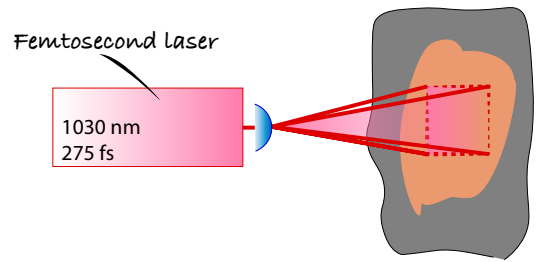

## 3- Analysis of the stone
Analysis of the stone by spectroscopy

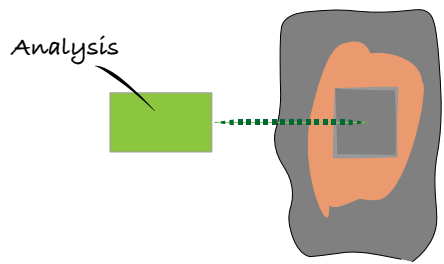

## 4- Identification of the stone
Comparison with reference spectra for identification

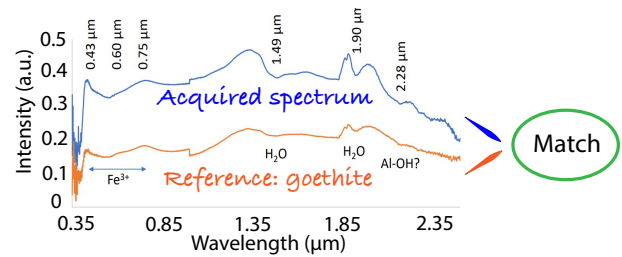

Fig. 2. Diagram of the process of using femtosecond pulse laser for mining planning: 1- application of slurry on the wall of a borehole for stabilisation, 2- opening a patch in the slurry by scanning focused femtosecond laser pulses, 3- analysis of the stone underneath, and 4- identification of the stone and minerals.

### *b. Femtosecond pulse laser system*

The experiments were performed with a Carbide CB3-40W femtosecond pulse laser (Light Conversion, Lithuania). This laser delivers pulses of 275 femtoseconds and has a maximum pulse energy of 0.4 mJ at a repetition rate of 100 kHz. The fundamental wavelength, infrared (IR) at 1030 nm, was used to investigate the ablation behaviour of the slurry and the different rocks. For beam scanning, a 10-facet polygon mirror (Precision Laser Scanning Inc., USA) (y-direction) and a galvanometer scanner (Cambridge Technology Inc., USA) (x-direction) were used. This allowed beam speeds of up to 880 m/s across a maximum scan area of 280 × 280 mm$^2$ at 52% duty cycle per polygon scan line for the laser, the beam speed being sufficiently high to operate in the single shot per spot regime, therefore avoiding possible heat accumulation effects. The laser pulses were focused on the sample with a quasi-telecentric f-Theta scanning lens of 540 mm working distance (S4LFT1420/449, Sill Optics GmbH, Germany). The focal spot size was reduced to 48.5 µm FWHM with a Gaussian intensity distribution using a 2X Achromatic Galilean Beam Expander (ThorLabs Inc. GBE02-C). The laser fluence was varied by attenuating the laser power. An air extraction system (KEMPER, fan capacity 1600 m$^3$/h, device capacity 950 m$^3$/h with exhaust arm) was located ~ 20 cm above the ablation area to collect the ablated particles and dust. Compressed air flow was directed through two nozzles with an exit opening diameter of 0.6 mm positioned at a distance of ~90 mm from the sample surface to lead the ablated particles into the extraction system and prevent dust from accumulating in front of the laser beam, which can reduce the interaction of the laser with the target surface. The ablation was undertaken in ambient laboratory conditions with a temperature of 23 ± 2ºC and an uncontrolled relative humidity typically in the 20-40 % range.

### c. *Slurry and rocks*

**Slurry:**
The samples are made from the material that constitutes the slurry of iron ore blast holes. The properties of the slurry can vary depending on its source, such as from the cone surface versus different depths within the same hole. Additionally, the slurry from one blast hole may differ significantly from that of another. Slurry characterisation is a topic that needs to be studied separately and is beyond the scope of this paper. The slurry sample used in this experiment was prepared by adding water to the mixture of dried material sampled from the surface of six different blast hole cones from multiple blast hole patterns of the same mine site (see Fig. 3).

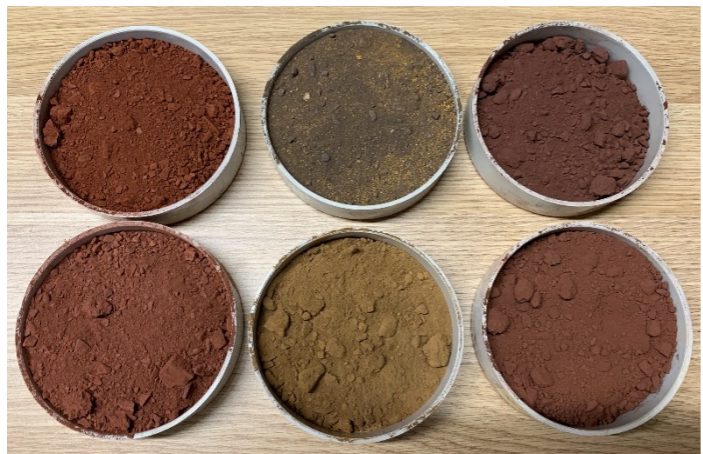

Fig. 3. The slurry sample was prepared by adding water to the mixture of dried material sampled from the surface of six different blast hole cones.

Water (between 2 and 3 mL) was added to a mixture of dried samples collected from the surface of six different blast hole cones (10g per sample) in small Petri dishes until a homogenous mixture was obtained. The samples were left overnight and lost some water. The samples were weighed a first time after mixing, and a second time just before the laser ablation. The water content was then estimated by calculating the ratio of water remaining in the sample (in g) to the amount of dry slurry (in g). The water content in the prepared samples varied from < 1 wt% to 14 wt%.

**Rocks:**
Five different rocks from a BIF (Banded Iron Formation) deposit of the Brockman Iron Formation in Western Australia were selected for the study. Each rock contained some silicas (such as microcrystalline quartz $SiO_2$) and clays (such as Kaolinite $Al_2Si_2O_5(OH)_4$) in addition to their predominant constituent, those rocks were:
- **Banded iron**: a sedimentary rock made of alternating layers of Iron oxides (generally hematite $Fe_2O_3$ or magnetite $Fe_3O_4$) and iron-poor chert (microcrystalline quartz $SiO_2$).
- **Limonite:** $FeO(OH) \cdot nH_2O$ hydrated ferric iron oxide.
- **Goethite** (vitreous) - α-$FeO(OH)$ containing ferric iron with notable pores.
- **Shale:** a sedimentary rock made of clay minerals (hydrous aluminium phyllosilicates such as kaolinite $Al_2Si_2O_5(OH)_4$) and fragments of other minerals such as quartz ($SiO_2$) and calcite ($CaCO_3$).
- **Hematite:** $Fe_2O_3$ containing ferric iron with macro and micro pores.

Rock samples were cut into flat billets to obtain reliable measurements with the laser and the optical profilometer and reduce the irregularities of the rocks interfering with the results, noting that even after preparation of the samples, goethite had very uneven surfaces with numerous holes and shale was extremely soft. The rocks are presented in Fig. 4 before sample preparation.

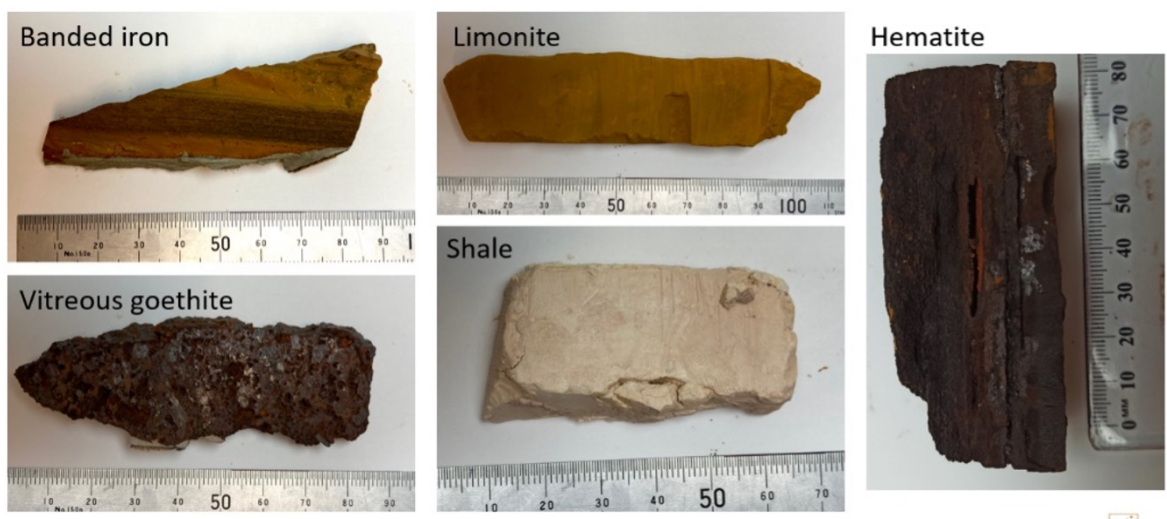

Fig. 4. Photography of banded iron, limonite, goethite, shale, and hematite samples.

### d. Analytical methods

Examining the rock surfaces and assessing any laser-induced changes was done using visual assessment, optical profilometry, Fourier Transform Infrared spectroscopy (FTIR), colourimetry, and VIS/NIR spectroscopy.

**Optical profilometry:** Surface texture, roughness, and grooves' depth were measured with an optical profilometer (Veeco Wyko NT9100, Bruker, USA) in vertical scanning interferometry mode (VSI) using a 5 × objective and 0.55 × field-of-view multiplier.

**Colourimetry:** Colour changes were assessed by colourimetry using a CM-2600d/2700d spectrophotometer (Konica Minolta, Japan) emitting with the D65 illuminant and an observer at 10 degrees through a 3 mm mask diameter, a specular component included. The results were expressed in the CIELab colour space, in which the three-dimensional coordinates of the CIELab colour space are obtained from the tri-stimulus colour matching functions. They can be expressed as (L*, a*, b*) in cartesian coordinates, or (L*, C*, h*) in polar coordinates. L* indicates lightness, a* the red/green coordinate, b* the yellow/blue coordinate. They can be represented by a 3D sphere, as illustrated in Fig. 5. For the polar coordinates, C* describes chroma (saturation) and h* is the hue angle.

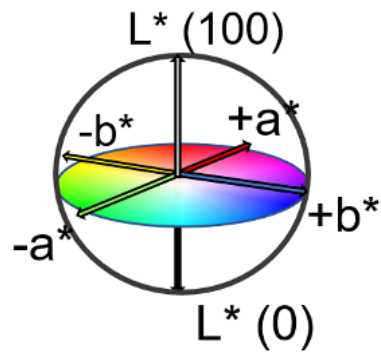

Fig. 5. CIELab colour space.

**Fourier Transform Infrared spectroscopy (FTIR):** FTIR was undertaken to analyse the effect of the laser on the surfaces. Measurements were carried out in Diffuse Reflectance FTIR mode (DRIFT) with a FlexScan 4200 handheld spectrometer (Agilent, USA) and spectra were collected over the full spectral range [4000 – 650] cm$^{-1}$, with a resolution of 4 cm$^{-1}$, and an accumulation of 64 scans. The raw spectra were baseline-corrected with BSpline interpolation, smoothed with a Savitzky-Golay filter of polynomial order 3 and 20 points, and Kubelka-Munk transformed. Because of the relatively large aperture of the instrument, regions of interest (ROI) on the samples need to be at least 10 $\times$ 10 mm$^2$ for accurate measurements. The samples were held in contact with the instrument for the duration of the acquisition, which may have disturbed the surfaces of limonite and shale.

**FieldSpec VIS/NIR spectrometer:** For Visible (VIS)/Near Infrared (NIR) spectroscopy, a halogen lamp was used to illuminate the samples. The 8˚ angle of incidence focus gun (probe) was used, and the configuration leading to the highest reflectance was investigated by varying the angle of the lamp on the sample, and the angle of the probe on the sample. The optimal conditions were found to be at 45˚ (probe and light source). For calibration of the FieldSpec 3, a white balance plate was used. 100 spectra were accumulated for white balance. Auto-calibration of the instrument was then selected. 50 spectra were accumulated on the samples during measurements. White balance was done between each sample (approximately every 10 minutes).

### e. *Ablation threshold and efficiencies measurement procedure.*

To determine the ablation threshold, rate, and efficiencies, grooves were ablated in the samples at varying fluences up to 15.00 ± 0.05 J/cm$^2$. The ablated depths of the grooves were measured by optical profilometry in different sections and averaged over at least 5 measurements. The depths per shot were calculated and used to estimate the ablation rates (mm$^3$/min) and efficiencies (mm$^3$/min/W).

## 3. Results

### a. *Ablation threshold, rate, efficiency, and effects on the slurry*

Sections of slurry prepared with different water content were ablated at varying fluences. The ablation threshold was found to be 0.35 ± 0.05 J/cm$^2$ for all tested samples. The water content did not influence the ablation threshold. Discolouration of the slurry surface occurred at fluence above 0.20 ± 0.05 J/cm$^2$. Fig. 6 presents two samples after laser ablation at different fluences.

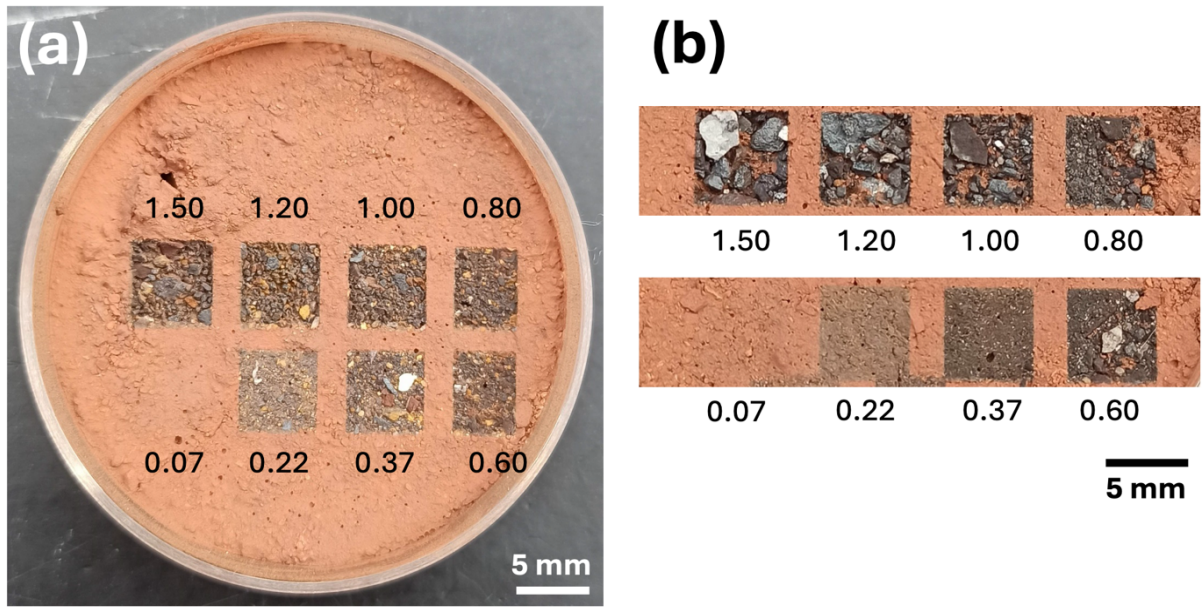

Fig. 6. Ablation of slurry with femtosecond laser pulses at 1030 nm. (a) and (b) represent the difference in the slurry composition with varying grain and pebble size. The numbers above or below the patches indicate laser fluence in J/cm$^2$.

Ablation efficiencies were determined for different water content: dry (< 1 wt% water content), 9%, 12%, and 14%, and the curves are presented in Fig. 7. All curves showed an increase in ablation efficiency up to 4.00 - 5.50 J/cm$^2$ where it reached a maximum. All samples tested with water content between 9% and 14% presented the same behaviour and characteristics, suggesting that the small variation in ablation efficiencies between water content resulted from sample-to-sample differences in the grain size, pebble content and size, and distribution through depth. The maximum ablation efficiency averaged around 10.6 mm$^3$/min/W. At higher fluences, the ablation efficiency decreased to stabilise around 2 - 4 mm$^3$/min/W.

However, dry slurry samples (i.e. very low water contents, < 1 wt%) exhibited remarkably different behaviour from those with higher amounts of water. The ablation efficiency maximum was almost two orders of magnitude higher than the maximum ablation efficiency achieved on the wet slurry, reaching a maximum of 482 mm$^3$/min/W at 5.3 J/cm$^2$ and reaching at least 440 mm$^3$/min/W for a fluence of 15 J/cm$^2$ (the highest tested fluence with the current experimental setup).

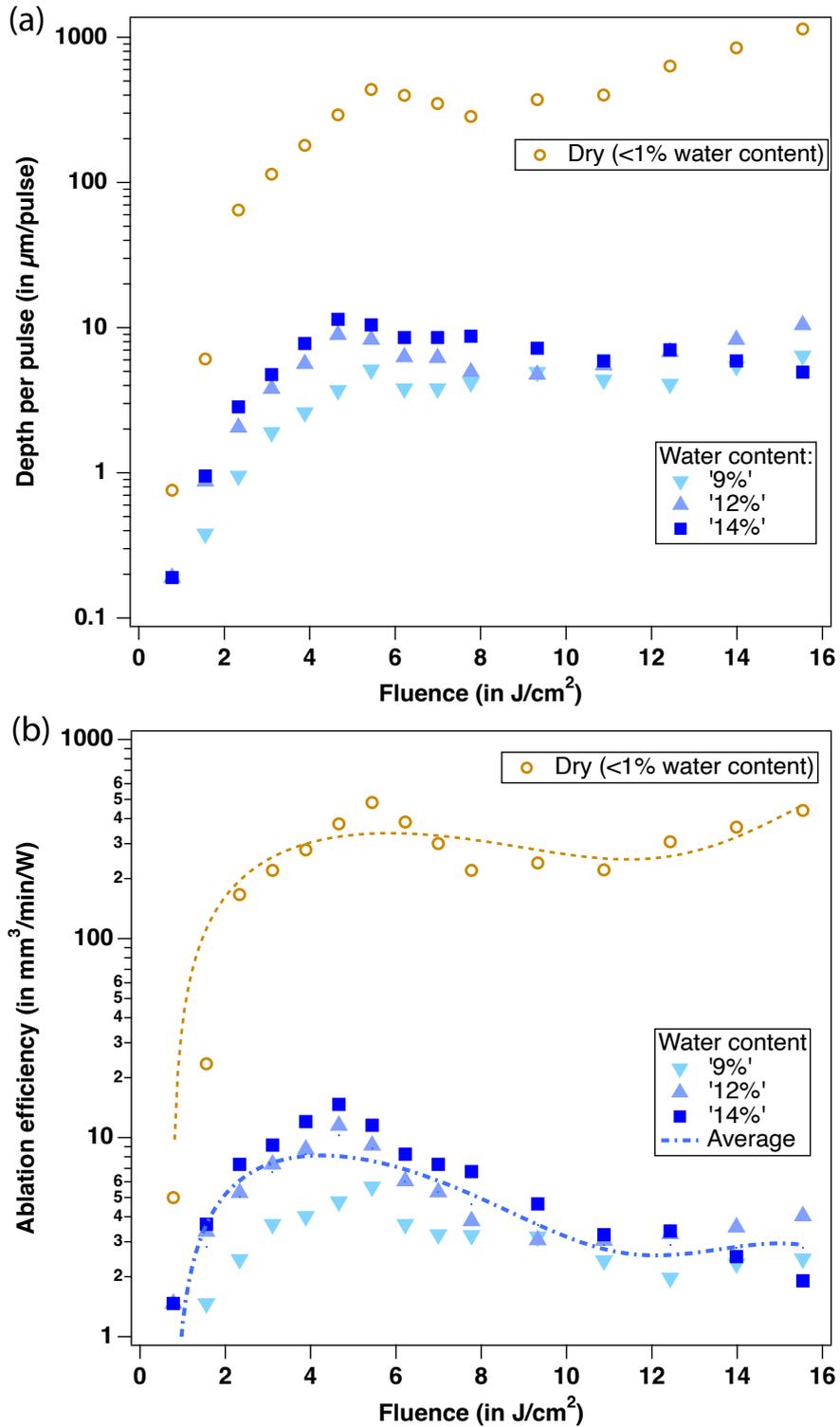

Fig. 7. Femtosecond pulse laser ablation of slurry at 1030 nm in function of the laser fluences for four different water content: dry (< 1 wt% water content), 9%, 12%, and 14%). (a) Depth per pulse and (b) ablation efficiencies, the brown dash line is a guide for the eye for the dry slurry curve, and the blue dotted line is the average for all water content.

The laser ablation results are summarised in Table 1 and highlight that independently of the water content the optimal ablation efficiency of the slurry is between 4.0 and 6.0 J/cm$^2$.

Table 1: Ablation efficiency of the slurry for different water content at 1030 nm wavelength using femtosecond laser pulses.

| Water content | < less than 1 % (dry) | 9% | 12% | 14% |
|---|---|---|---|---|
| Discolouration threshold, in J/cm$^2$ | 0.20 ± 0.05 | 0.20 ± 0.05 | 0.20 ± 0.05 | 0.20 ± 0.05 |
| Ablation threshold, in J/cm$^2$ | 0.35 ± 0.05 | 0.35 ± 0.05 | 0.35 ± 0.05 | 0.35 ± 0.05 |
| Maximum ablation efficiency, in mm$^3$/min/W | 482 at 5.30 J/cm$^2$ | 5.6 at 5.30 J/cm$^2$ | 11.5 at 4.70 J/cm$^2$ | 14.7 at 4.70 J/cm$^2$ |
|  |  | Average ablation efficiency with Water content: 10.6 mm$^3$/min/W | | |

These very high ablation efficiencies were achieved when grain size conditions were met. The samples contained many pebbles (of size >0.5 mm) which did not ablate as fast as the powdery iron particles. They were nonetheless removed by two mechanisms. Firstly, for samples standing vertically (as the bore wall would be), a combination of gravity and shock waves from the laser pulses led to the pebbles falling out of the matrix. Secondly, due to the large volume of dust created at such high efficiencies, an air jet was used to blow the dust away from the optics and the ablation area and into the extraction system located above the samples (about 20 cm). The air turbulence created in such a way assisted the removal of the pebbles, dropping from the matrix by aerodynamic drag forces once they were held only at the back. It is important to note that the air jet alone did not remove the slurry, as the sides of the ablated window were not disturbed (see Fig. 8). A 2.3 × 3 cm$^2$ window was ablated in the slurry.

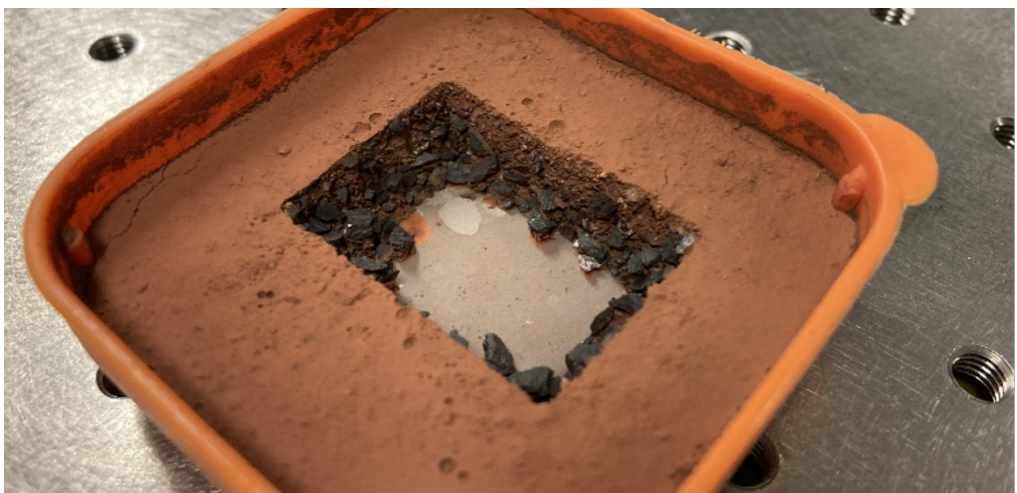

Fig. 8. Femtosecond pulse laser ablation of 1.3 cm deep slurry at 1030 nm. The size of the window is 2.3 × 3 cm$^2$. Discoloured pebbles are visible around the edges in black.

The 1.3 cm thick slurry was completely removed after 40 laser scans. For such thicknesses, the depth of focus (the Rayleigh range) of the laser spot must be considered to maintain the highest ablation rate. Briefly, if the slurry is ablated beyond the Rayleigh range, the beam starts diverging, resulting in a larger beam diameter at the ablation point compared to the beam diameter at the focal plane. This increased beam diameter can lead to a reduced laser fluence and lower ablation efficiency. Consequently, it might be necessary to refocus the laser during the ablation.

### b. *Femtosecond laser ablation of stones*

#### i. Ablation threshold and efficiencies

The ablation threshold and efficiency of the stones were investigated by irradiating square zones of 5 × 5 mm$^2$ on the surface at increasing laser fluence and the results are summarised in Table 2. The ablation threshold was determined to be 0.30 - 0.40 ± 0.05 J/cm$^2$ for all stones. Discolouration appeared at lower fluences, below their ablation threshold, between 0.07 ± 0.05 J/cm$^2$ for limonite, and 0.15 ± 0.05 for hematite and shale. Goethite and banded iron had a slightly higher discolouration threshold at 0.30 ± 0.05 J/cm$^2$.

Table 2: Ablation threshold and ablation efficiencies of hematite, goethite, banded iron, shale and limonite at 1030 nm using femtosecond laser pulses.

|  | Hematite | Goethite | Banded Iron | Shale | Limonite |
|---|---|---|---|---|---|
| Discolouration threshold, in J/cm$^2$ | 0.15 ± 0.05 | 0.30 ± 0.05 | 0.30 ± 0.05 | 0.15 ± 0.05 | 0.07 ± 0.05 |
| Ablation threshold, in J/cm$^2$ | 0.30 - 0.40 ± 0.05 | 0.30 - 0.40 ± 0.05 | 0.30 - 0.40 ± 0.05 | 0.30 - 0.40 ± 0.05 | 0.30 - 0.40 ± 0.05 |
| Ablation efficiency, in mm$^3$/min/W | 1.0 - 1.5 at 4.00 – 6.00 J/cm$^2$ | 1.0 - 1.5 at 4.00 - 6.00 J/cm$^2$ | 1.0 - 1.5 at 4.00- 6.00 J/cm$^2$ | 2.9 * at 15.50 J/cm$^2$ | 15.6 at 7.70 J/cm$^2$ |

*Maximum not reached, higher fluence needs to be investigated.

The discolouration observed on the surface appeared shallow and occurred only at a depth of a few µm. One possible explanation could be a change in oxidation states of iron or charge transfers, that can happen when rocks are irradiated at wavelengths between ~330 and ~2500 nm. No melting of the rocks was noted with digital microscopy for all tested fluences. Fig. 9 presents the laser-processed zones of each rock for different laser fluences.

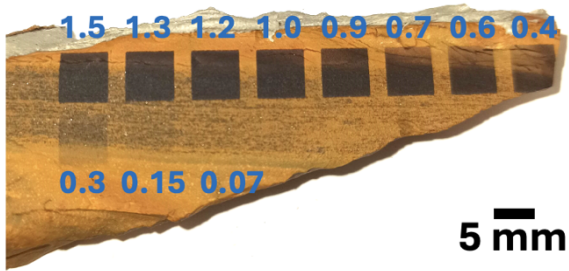
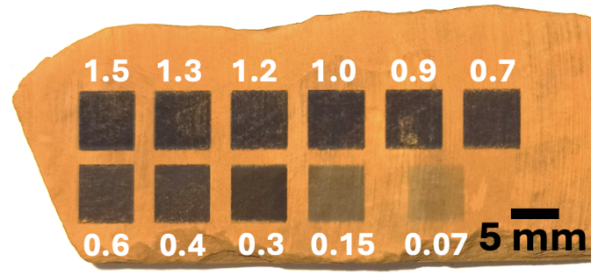
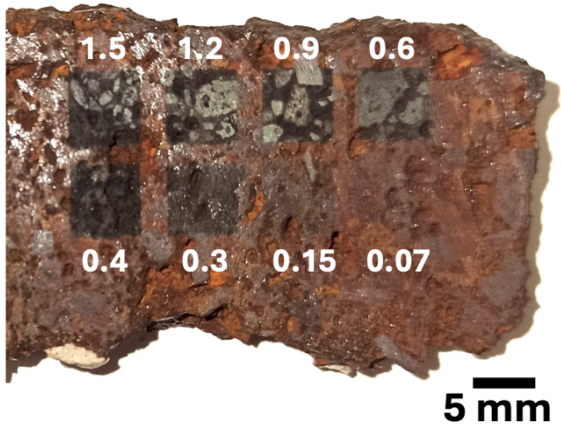
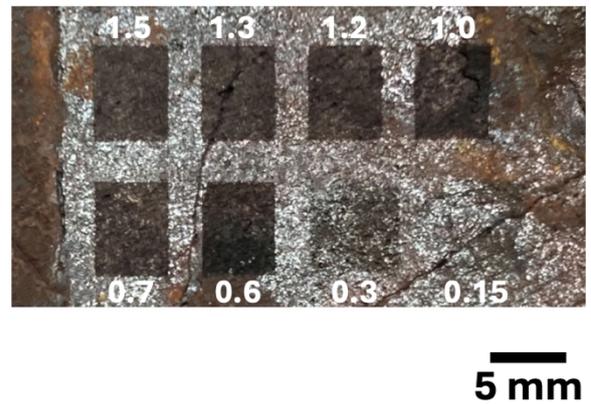
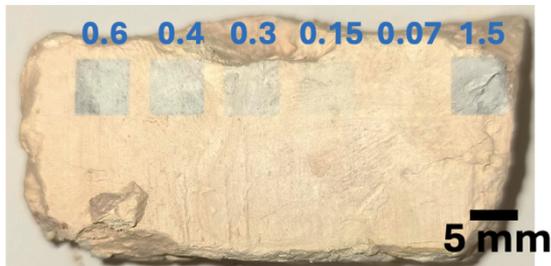

Fig. 9. Femtosecond pulse laser ablation of the stones at 1030 nm: (a) banded iron; (b) limonite; (c) goethite; (d) hematite; (e) shale. Numbers adjacent to the irradiated square are the laser fluence in J/cm$^2$.

The ablation efficiency was measured for all stones and the results presented in Fig. 10. Banded iron, hematite, and goethite showed similar behaviour: an increase in ablation efficiency with the increase of the laser fluence until reaching a maximum of 1.0 - 1.5 mm$^3$/min/W at fluences around 4.00 - 6.00 ± 0.05 J/cm$^2$. At higher fluences, the ablation efficiency decreased. Shale showed a steady increase in ablation efficiency without reaching a maximum (in the range of tested fluences). It is suspected that its maximum is located at higher fluences. The variation in ablation rate between shale and the other stones can be explained by their difference in composition. The ablation of limonite was much more efficient than the other stones; grooves were much deeper for the same number of laser pulses, and a maximum ablation of 15.6 mm$^3$/min/W was reached at 7.70 ± 0.05 J/cm$^2$.

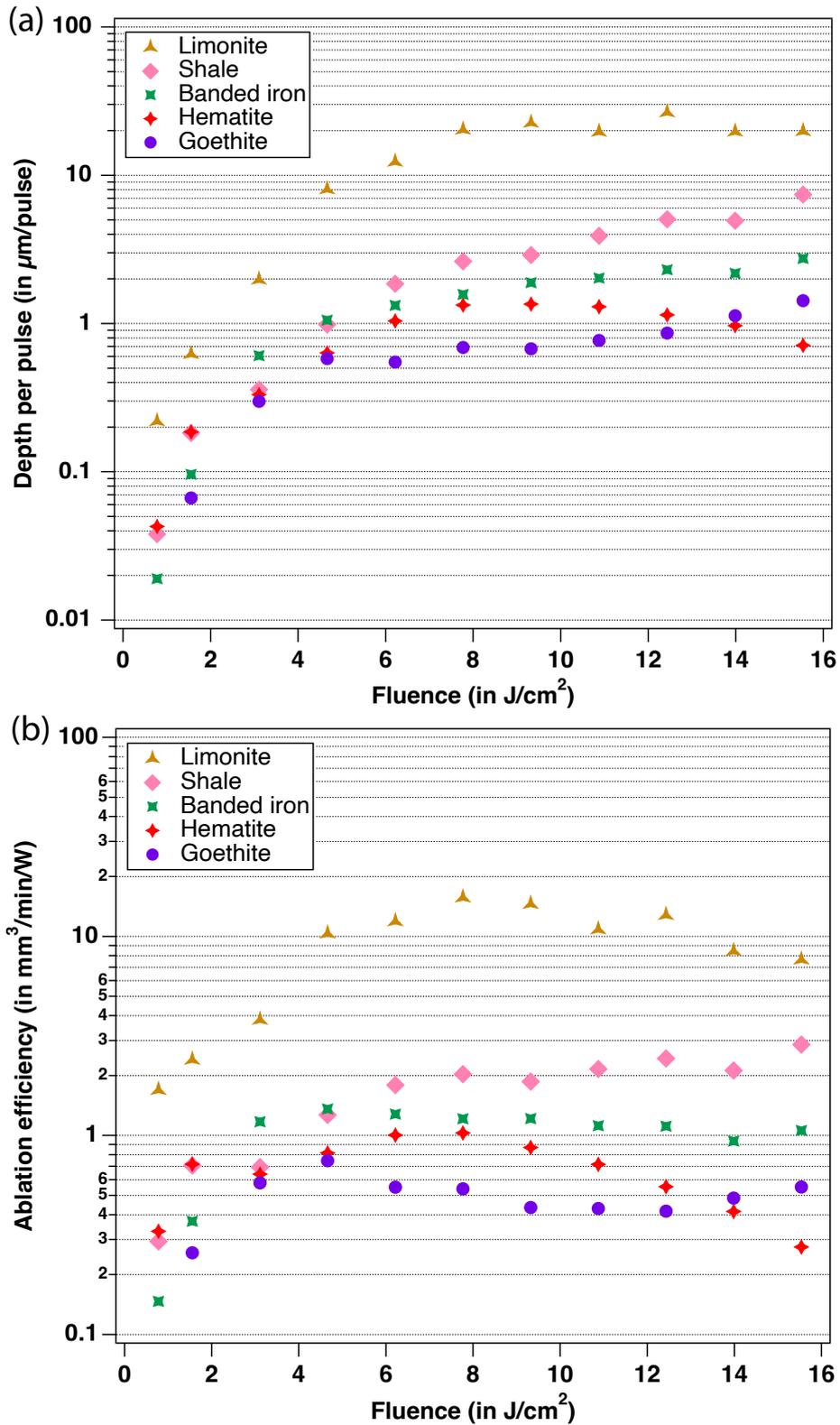

Fig. 10. Femtosecond pulse laser ablation of hematite, goethite, banded iron, shale and limonite as a function of the laser fluences: (a) depth per pulse and (b) ablation efficiencies.

### c. Analysis of stone surface
#### i. Colourimetry

Colourimetry measurements were undertaken on all the rocks before and after laser irradiation and the results are summarised in Table 3.

**Banded iron and limonite:** before laser irradiation, banded iron and limonite exhibited high b* values (characteristic yellow colour): 21.7 units for banded iron and 43.4 units for limonite. After laser irradiation at the fluence of 0.80 ± 0.05 J/cm$^2$, the b* parameter decreased by 91% (to 1.9 units) for banded iron, and 83% (to 7.2 units) for limonite, which indicates the loss of most of the surface colouration. After irradiation at 1.50 ± 0.05 J/cm$^2$, banded iron lost 97% of its colour and 96% for limonite. The L* parameter was severely affected too, dropping from 46.6 to 29.6 units for banded iron and from 52.2 to 21.9 units for limonite, which indicates a severely darker surface colour. The a* parameter, characteristic of the red component, was reduced (92% for banded iron and 84.6% for limonite). The discolouration is thus clearly observed on both rocks.

**Goethite and hematite:** for goethite and hematite, the impact of the laser was not as significant, mainly because the stones were originally darker, but the discolouration was still visible. The L* parameter decreased slightly, from 33.2 to 28.8 units for goethite, and 27.0 to 25.4 units for hematite. Both a* and b* parameters were affected by the laser, which confirmed the loss of surface colour. For goethite, a* decreased by 82%, and b* by 86%; for hematite, a* decreased by 105% (going into negative values) and b* by -112.5% (also negative).

**Shale:** shale discoloured to a greyish tone in the laser-processed area, which is confirmed by the L* parameter slightly decreasing from 85.6 units before laser irradiation to 79.1 units after irradiation at 1.50 ± 0.05 J/cm$^2$. However, the a* and b* parameters were severely affected, with a* decreasing by 106% and b* by 96%.

Table 3. Colourimetry results in the function of L, a*, and b* parameters for the stone before and after femtosecond laser irradiation at 1.50 ± 0.05 J/cm$^2$. For all the stones, a general trend is noted, the increase of discolouration with the increase of laser fluence.

|  |  | L* | a* | b* |
|---|---|---|---|---|
| Banded iron | before (reference) | 46.5 | 11.2 | 21.7 |
|  | 1.50 ± 0.05 J/cm$^2$ | 29.6 | 0.9 | 0.6 |
| Limonite | before (reference) | 52.2 | 17.6 | 43.4 |
|  | 1.50 ± 0.05 J/cm$^2$ | 21.9 | 2.7 | 1.7 |
| Goethite | before (reference) | 33.2 | 5.9 | 5.1 |
|  | 1.50 ± 0.05 J/cm$^2$ | 28.8 | 1.0 | 0.7 |
| Hematite | before (reference) | 27.0 | 3.3 | 2.0 |
|  | 1.50 ± 0.05 J/cm$^2$ | 25.4 | -0.2 | -0.3 |
| Shale | before (reference) | 85.6 | 4.7 | 12.5 |
|  | 1.50 ± 0.05 J/cm$^2$ | 79.1 | -0.3 | 0.5 |

To summarise, colourimetry showed an intense loss of colour for all stones irradiated with the laser, which is consistent with observations with the naked eye. Banded iron, limonite, goethite and hematite turned black, whereas shale turned grey.

### ii. Fourier Transform Infrared spectroscopy (FTIR)

FTIR analysis was undertaken on all stones after laser irradiation. For all the rocks, it was noted that the intensity of the signal decreases as the fluence increases, which can indicate surface modifications.

**Limonite:** limonite presented a main feature at 3400 $cm^{-1}$, characteristic of water OH stretching vibration [29], this feature decreased in intensity as the fluence increased. This could be interpreted as the OH groups being removed during the laser irradiation and could correlate to oxidation.

**Shale:** shale exhibits characteristic bands of clay, with the main features at 3697 $cm^{-1}$, 3660 $cm^{-1}$ and 3627 $cm^{-1}$, attributed to kaolinite [30]. Two $SiO_2$ stretching vibrations were visible at 1126 and 1056 $cm^{-1}$, and an additional OH vibration band at 1010 $cm^{-1}$. Al-OH deformation was visible at 910 $cm^{-1}$ [30]. A band at 1640 $cm^{-1}$ can be attributed to the bending of HOH bonds. For shale as well, the main difference is a decreasing intensity of the water band at 3400 $cm^{-1}$ with the increase of laser fluence and a decreasing intensity of the SiO band at 1056 $cm^{-1}$. Although decreasing, the latter does not exhibit any shifts, loss of feature (flatline) or broadenings that could indicate damage to the structure. However, a possible reason is that DRIFT may go through the damage layer instead of measuring it. FTIR in Attenuated Total Reflectance (ATR) may provide more insightful information.

**Hematite:** Hematite consists of different components, making the region [3000-1000] $cm^{-1}$ complex to interpret. The main common feature of all spectra is a band at 700 $cm^{-1}$ which could be attributed to FeO stretching vibrations. This band does not change depending on the laser fluence. The observations as stated above for shale and limonite were made for the OH band at 3400 $cm^{-1}$, the band intensity is decreasing with the increase of laser fluence.

**Goethite:** the spectra show the FeO band at 700 $cm^{-1}$, a shallow OH band at 3400 $cm^{-1}$, and an overtone or combination peak at 2400 $cm^{-1}$. As previously mentioned, the intensity of the OH band at 3400 $cm^{-1}$ is decreasing with increased fluence. The position and the intensity of the FeO peak did not change during laser irradiation.

**Banded iron:** the spectra show features characteristic of quartz ($SiO_2$) at 1214 and 1085 $cm^{-1}$ (stretching vibration SiO). Additional silicate bands are present at 804 and 780 $cm^{-1}$. The bands in the 2000-1400 $cm^{-1}$ range are likely to be artefacts. As previously mentioned, the intensity of the OH band at 3400 $cm^{-1}$ decreased with increased fluence. The SiO bands did not change with the fluence (a lower intensity is observed but no broadenings, loss of features, or shifts).

To summarise, for all the tested stones, changes occurred to the OH bands located at 3400 $cm^{-1}$ during laser irradiation, which could be attributed to a loss of water or molecular OH. No shifts, broadenings or loss of features were observed, and the stones were still clearly identifiable despite their colour change.

### iii. VIS/NIR spectrometry

Visible (VIS) / Near Infrared (NIR) spectroscopy was undertaken on the rocks, as this method is commonly used for diagnostics in the mining sector. In the VIS/NIR domain, ochreous and vitreous goethite typically show an absorption shoulder at 0.52 µm, 0.70 µm [31]. The major diagnostic feature for ferric oxide can be observed at 0.90 µm and correlates to the hematite/goethite ratio. In the spectra collected on the rocks, this band at 0.90 µm was weak if observed at all, but a strong band was observed around 0.85 µm. A steeper reflectance ramp between 1.00 and 1.40 µm is typical of vitreous goethite. The spectrum of goethite typically includes Fe electronic bands near 0.64–0.65 and 0.92–0.93 µm, a Fe-OH doublet near 2.4 and 2.5 µm, a broad OH/$H_2O$ band centred near 3.1–3.2 µm [32], and a weak band near 1.9 µm due to $H_2O$.

The spectra of goethite here mainly showed Fe electronic bands at 0.42 µm, 0.60 µm, and 0.77 µm. A decrease in S/N was noted after laser irradiation. Spectra of fine-grained (~10 – 125 µm grain size) hematite exhibit a band at 0.53 µm, a shoulder near 0.6 µm, and a band near 0.85 µm that shifts toward 0.88 µm [32]. In contrast, coarse-grained, grey hematite is spectrally dark and nearly featureless in the VIS/NIR region. The main difference between non-irradiated and irradiated goethite was a change in slope for the 0.6 µm band, possibly due to an electronic transition of the $Fe^{3+}$ ion. The effect is less important at lower fluences and similar behaviour was observed on the limonite and banded iron spectra.

The characteristic features in the SWIR/MIR region of clay minerals are a doublet at 2.20 µm associated with a second peak at 2.16 or 2.18 µm, attributed to stretching vibration of the inner hydroxyl groups (hydroxyl groups bond to the tetrahedral layer) and outer hydroxyl group ($\nu + \delta Al_2OH_0$) [31]. Changes to this doublet represent changes in the structure and crystallinity of the kaolin group. Here, no changes were noted on the hydroxyl peaks: no shifts, broadenings, or loss of features, indicating no effect on the structure and crystallinity of the kaolin group. A sharp pair of bands due to the Al–OH stretching overtone are observed at 1.39 and 1.41 µm [31, 32], and no particular changes were noted on those bands either. The main difference can be seen on the slope leading to the 0.59 µm band, which is affected by the laser and increases as the fluence increases, possibly indicating an electronic transition of $Fe^{3+}$ ions.

To summarise, the main effect noted on all the rocks after irradiation by the laser using VIS/NIR spectroscopy is the change in the slope of the $Fe^{3+}$ electronic transition band around 0.60 µm, suggesting a possible change in the oxidation state of iron, which could be better observed with X-ray Photoelectron Spectroscopy (XPS).

### 4. Femtosecond laser removal of slurry from goethite

For the demonstration of slurry removal, the goethite sample was selected and covered by a ~1.5 cm thick layer of slurry. The sample was left to dry until a content of 3% water was obtained. The sample before laser irradiation is presented in Fig. 11a. Ablation of a 2.5 × 3 cm² window at a laser fluence of 5.40 ± 0.05 J/cm² was then undertaken until total removal of the slurry layer was reached. The laser was scanned from left to right 100 times (100 scans). Fig. 11b presents the goethite sample after the laser irradiation. All the slurry was removed from the targeted area and the goethite was exposed for analysis. Since typically the stones and slurry share the same rock composition, it becomes impossible to prevent ablation of the stone once it is exposed to the laser beam while the slurry is not yet fully removed. Therefore, it is necessary to ablate a small thickness of the stone to ensure complete removal of the slurry. Discolouration, consistent with the observations made in the previous sections, can be seen in the laser-processed area, but no morphological damage to the stone was noted (no visible change in roughness, no apparent melting). FTIR and VIS/NIR spectroscopy revealed the typical features of goethite together with the decrease in intensity of the OH band at 3400 cm$^{-1}$ and a change in the slope of the $Fe^{3+}$ electronic transition band around 0.60 μm, as described in section 4. The identification of goethite was clear when compared to the reference spectrum of goethite.

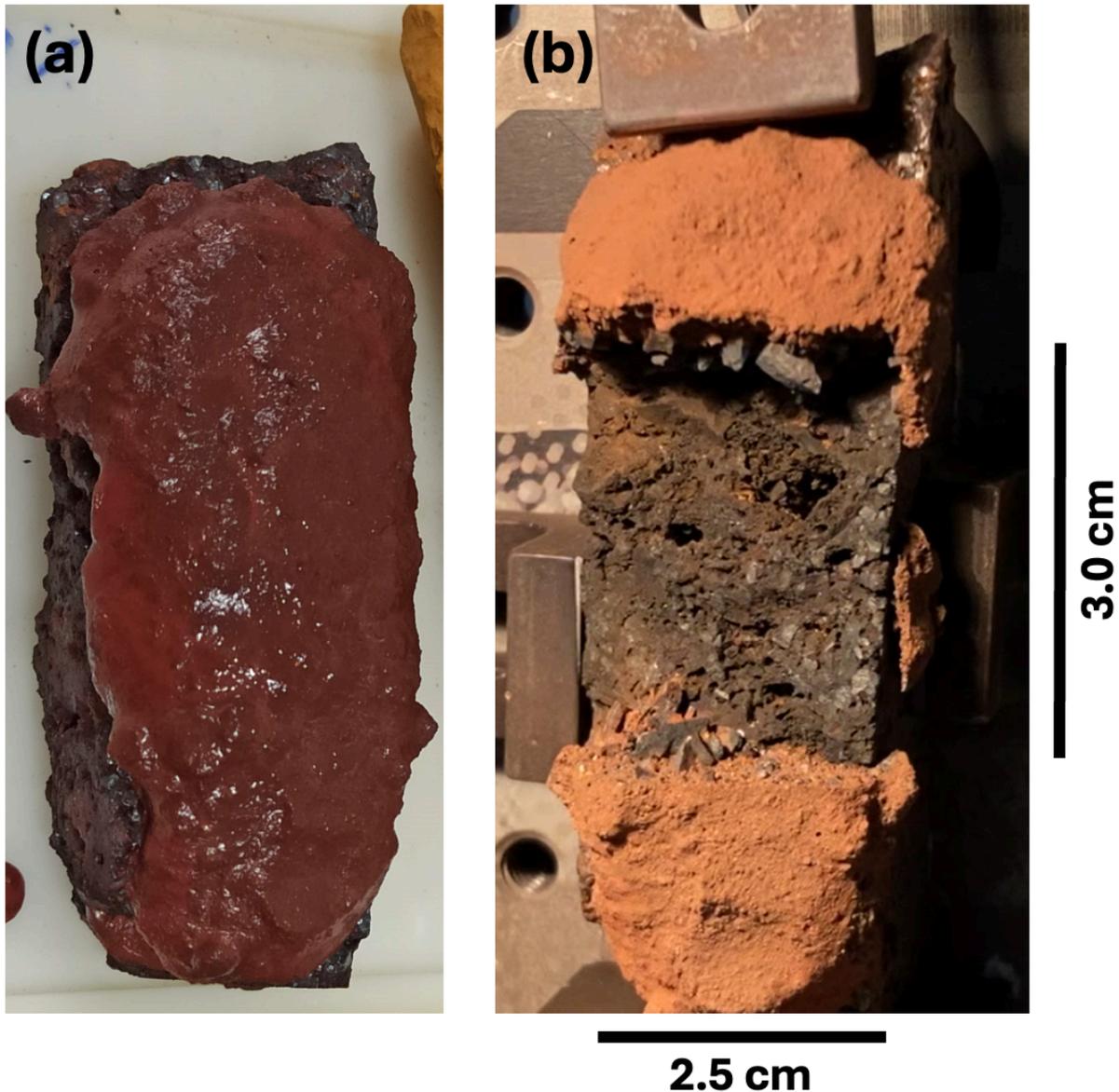

Fig. 11. (a) Goethite sample covered with 1.5 cm thick slurry with 3% water content. (b) Femtosecond pulse laser ablation of the slurry over a 2.5 × 3 cm$^2$ central area on the sample until the slurry is completely removed and the stone underneath exposed.

5.  Conclusion

**Femtosecond laser removal of slurry:** Precise and fast ablation of wet and dry slurry was achieved using femtosecond laser pulses and high ablation rates were obtained: 482 mm$^3$/min/W at 5.30 ± 0.05 J/cm$^2$ for dry (less than 1% of water content) and an average of 10.6 mm$^3$/min/W for wet slurry (slurry with varying water content) for fluences in the range of 4.00 to 5.50 ± 0.05 J/cm$^2$. The ablation threshold of the slurry, independent of the water content, was found to be at 0.35 ± 0.05 J/cm$^2$. Removal of slurry in containers and on the stone was found very effective. A slurry layer up to 1.5 cm thick was ablated, and it would be possible to remove even thicker layers by carefully considering the depth of focus (the Rayleigh range) the laser spot and the focusing conditions to maintain the highest ablation rate. This can be done by moving the laser focusing elements on the targeted area with a controlled, possibly motorised, stage. The ablation process was fast, and the airflow and extraction system helped dislodge the small rocks that became loose during the ablation process.

**Femtosecond laser ablation of stones:** The ablation threshold was determined to be 0.30 ± 0.05 J/cm$^2$, with discolouration occurring at slightly lower fluence. Discolouration could not be avoided using the 1030 nm wavelength. However, no significant damage was observed, and only the surface layer of the stones was affected. The discolouration is likely attributed to a change in the oxidation state of iron, as suggested by the VNIR spectroscopy results. To minimise discolouration, it would be advisable to use laser wavelengths in the mid-IR domain, where phenomena such as charge transfer and crystal field transitions are less intense.

**General conclusion:**
Using a femtosecond laser to ablate slurry sections offers significant advantages for the subsequent identification and analysis of the rock underneath. While there may be some challenges due to loss of surface colour over a superficial layer, advanced techniques such as FTIR and VIS/NIR spectroscopy can still accurately identify the rock. A potential direction of research would be to study the femtosecond laser ablation of the rocks and slurry in the mid-infrared wavelength domain, as this wavelength range can potentially reduce the risk of altering the stone's visible colour and ensures that crucial spectral features needed for identification remain intact.

The slurry was efficiently removed using the femtosecond laser pulses, offering a valuable solution for detailed mining operations and planning. This method enables precise material removal at the micrometric scale with minimal damage, which is essential for identifying high-value minerals. This precision helps optimise resource extraction, making them invaluable tools in mining.

**Declaration of Competing Interest:**
The authors declare that they have no known competing financial interests or personal relationships that could have appeared to influence the work reported in this paper.

**Data availability:**
Data will be made available on request.

**Acknowledgments:**
The authors would like to thank Michael Turner from the Research School of Physics of the Australian National University for the sample preparation, and the University of Canberra for access to the colorimeter and FTIR spectrometer. This work has been supported by the Australian Centre for Robotics and the Rio Tinto Centre for Mine Automation.

**Funding sources:**
This work has been supported by the Australian Centre for Robotics and the Rio Tinto Centre for Mine Automation.

**CRediT author statement:**
**Julia Brand:** Writing - Original Draft, Investigation, Conceptualization, Formal analysis, Visualization. **Ksenia Maximova:** Writing - Review & Editing, Investigation, Formal analysis, Visualization. **Steve Madden:** Software, Conceptualization, Methodology. **Andrei V. Rode:** Writing - Review & Editing, Conceptualization, Methodology. **Ehsan Mihankhah:** Writing - Review & Editing, Conceptualization, Methodology. **John Zigman:** Writing - Review & Editing, Conceptualization, Methodology. **Andrew J. Hill:** Conceptualization, Supervision, Project administration, Writing – Review & Editing, Funding acquisition. **Ludovic Rapp:** Conceptualization, Methodology, Validation, Investigation, Resources, Formal analysis, Visualization, Writing – Original draft, Writing - Review & Editing, Supervision, Project administration, Funding acquisition.

**References**

[1] Summerfield, D. 2020. Australian Resource Reviews: Iron Ore 2019. Geoscience Australia, Canberra. http://dx.doi.org/10.11636/9781925848670

[2] Unknown Author, May 2018. Fact sheet on: Iron. Geoscience Australia, Canberra. Accessible online: https://www.ga.gov.au/education/classroom-resources/minerals-energy/australian-mineral-facts/iron

[3] R. Leung, A. Lowe, A. Chlingaryan, A. Melkumyan, and J. Zigman, *Bayesian surface warping approach for rectifying geological boundaries using displacement likelihood and evidence from geochemical assays*, ACM Transactions on Spatial Algorithms and Systems, vol. 8, no. 1, pp. 1–23, Mar. 2022.

[4] A. Ball, J. Zigman, A. Melkumyan, A. Chlingaryan, K. Silversides, and R, Leung, *Addressing Application Challenges with Large-Scale Geological Boundary Modelling*, Springer International Publishing, Geostatistics Toronto 2021, pp 221–236, 2023.

[5] S. Talesh Hosseini, O. Asghari, J. Benndorf, and X. Emery, *Real-time uncertain geological boundaries updating for improved block model quality control based on blast hole data: A case study for Golgohar iron ore mine in Southeastern Iran*, Mathematical Geosciences, vol. 55, no. 4, pp. 541–562, 2023.

[6] N. Ahsan, S. Scheding, S. T. Monteiro, R. Leung, C. McHugh, and D. Robinson, *Adaptive sampling applied to blast-hole drilling in surface mining*, International Journal of Rock Mechanics and Mining Sciences, vol. 75, pp. 244–255, 2015.

[7] R. Leung, A. J. Hill and A. Melkumyan, "Automation and Artificial Intelligence Technology in Surface Mining: A Brief Introduction to Open-Pit Operations in the Pilbara," in IEEE Robotics & Automation Magazine, doi: 10.1109/MRA.2023.3328457.

[8] L. Liu, E. Mihankhah, and A. Hill. "Robust blasthole detection for a mine-site inspection robot." (ACRA 2022), [Online]. Available: https://ssl.linklings.net/conferences/ acra/acra2022_proceedings/views/includes/files/ pap134s2.pdf.

[9] J. Brand, A. V. Rode, S. Madden, A. Wain, P. L. King and L. Rapp, *Ultrashort pulsed laser ablation of granite for stone conservation,* Optics and Laser Technology 151 (2022) 108057

[10] V. Zinnecker, S. Madden. C. Stokes-Griffin, P. Compston, A. V. Rode and L. Rapp, *Ultrashort pulse laser ablation of steel in air ambient,* Optics and Laser Technology (2022) 148 107757

[11] Gamaly, E.G., A. V. Rode, B. Luther-Davies and V. T. Tikhonchuk, *Ablation of solids by femtosecond lasers: Ablation mechanism and ablation thresholds for metals and dielectrics.* Physics of Plasmas, 2002. **9**(3): p. 949-957.

[12] J. Brand, A. Wain, A. V. Rode, S. Madden and L. Rapp, *Towards safe and effective femtosecond laser cleaning for the preservation of historic monuments,* Applied Physics A 129:246 (2023)

[13] J. Brand, A. Wain, A. V. Rode, S. Madden, P. King and L. Rapp, *Femtosecond pulse laser cleaning of spray paints from heritage stone surfaces,* Optics Express 30 17 (2022) 31122

[14] J. S. Pozo-Antonio, A. Papanikolaou, K. Melessanaki, T. Rivas and P. Pouli, *Laser-Assisted Removal of Graffiti from Granite: Advantages of the Simultaneous Use of Two Wavelengths.* Coatings, 2018. **8**(4): p. 124-144.

[15] J. S. Pozo-Antonio, T. Rivas, M.P. Fiorucci, A. J. López and A. Ramil, *Effectiveness and harmfulness evaluation of graffiti cleaning by mechanical, chemical and laser procedures on granite.* Microchemical Journal, 2016. **125**: p. 1-9.

[16] J. Brand, A. Wain, A. V. Rode, S. Madden, P. King and L. Rapp, *Comparison between nanosecond and femtosecond laser pulses for the removal of spray paint from granite surfaces,* Journal of Cultural Heritage 62 329-338 (2023)

[17] J. Brand, S. Madden, A. Wain, A. V. Rode and L. Rapp, *Femtosecond pulse laser cleaning of marble,* Applied Surface Science (2023)

[18] J. Brand, A. V. Rode, S. Madden, A. Wain and L. Rapp, *Femtosecond pulse laser cleaning of Makrana marble and semi-precious stones for the preservation of the Holy Samadh*, International Association for Bridge and Structural Engineering (IABSE) Conference proceedings, ISBN:978-3-85748-190-1 (2023)

[19] J. Brand, A. Wain, A. V. Rode, S. Madden, P. King, M. Mohan, W. Kaluarachchi, J. Ratnayake and L. Rapp, *Femtosecond pulse laser cleaning biofilm and dirt: preserving the Sydney Harbour Bridge,* Journal of Cultural Heritage 60 86-94 (2023)

[20] A. J. López, T. Rivas, J. Lamas, A. Ramil and A. Yáñez, *Optimisation of laser removal of biological crusts in granites.* Applied Physics A, 2010. **100**(3): p. 733-739.

[21] J. S. Pozo-Antonio, P. Barreiro, G. Paz-Bermúdez, P. González and A. B. Fernandes, *Effectiveness and durability of chemical- and laser-based cleanings of lichen mosaics on schists at archaeological sites.* International Biodeterioration & Biodegradation, 2021. **163**: p. 105276.

[22] J. S. Pozo-Antonio, M.P. Fiorucci, A. Ramil, A. J. López and T. Rivas, Evaluation of the effectiveness of laser crust removal on granites by means of hyperspectral imaging techniques. Applied Surface Science, 2015. 347: p. 832-838.

[23] T. Rivas, J. S. Pozo-Antonio, M. E. López de Silanes, A. Ramil and A. J. López, Laser versus scalpel cleaning of crustose lichens on granite. Applied Surface Science, 2018. 440: p. 467-476.

[24] S. Pozo, P. Barreiro, T. Rivas, P. González and M. P. Fiorucci, Effectiveness and harmful effects of removal sulphated black crust from granite using Nd:YAG nanosecond pulsed laser. Applied Surface Science, 2014. 302: p. 309-313.

[25] J. S. Pozo-Antonio, A. Papanikolaou, A. Philippidis, K. Melessanaki, T. Rivas and P. Pouli, *Cleaning of gypsum-rich black crusts on granite using a dual wavelength Q-Switched Nd:YAG laser.* Construction and Building Materials, 2019. **226**: p. 721-733.

[26] J. S. Pozo-Antonio, A. Ramil, T. Rivas, A. J. López and M. P. Fiorucci, *Effectiveness of chemical, mechanical and laser cleaning methods of sulphated black crusts developed on granite.* Construction and Building Materials, 2016. **112**: p. 682-690.

[27] C. Rodriguez-Navarro, K. Elert, E. Sebastián, R. Esbert, C. Grossi, A. Rojo, Fco. J. Alonso, M. Montoto and J. Ordaz, "Laser cleaning of stone materials: an overview of current research," Studies in Conservation, vol. 48, no. sup1, pp. 65-82, 2003.

[28] T. Rivas, A. J. Lopez, A. Ramil, S. Pozo, M. P. Fiorucci, M. E. López de Silanes, A. García, J. R. Vazquez de Aldana, C. Romero and P. Moreno, Comparative study of ornamental granite cleaning using femtosecond and nanosecond pulsed lasers. Applied Surface Science, 2013. 278: p. 226-233.

[29] G. Socrates, Infrared and Raman characteristic group frequencies, 3rd Edition. 2001.

[30] M. Labus and M. Lempart, "*Studies of Polish Paleozoic shale rocks using FTIR and TG/DSC methods*," Journal of Petroleum Science and Engineering, vol. 161, pp. 311-318, 2018/02/01/ 2018.

[31] Ramanaidou, E. (2015). *Characterization of iron ore by visible and infrared reflectance and, Raman spectroscopies*. Iron Ore, 191–228.

[32] Bishop, J. (2019). *Visible and Near-Infrared Reflectance Spectroscopy: Laboratory Spectra of Geologic Materials*. In J. Bishop, J. Bell III, & J. Moersch (Eds.), Remote Compositional Analysis: Techniques for

Understanding Spectroscopy, Mineralogy, and Geochemistry of Planetary Surfaces (Cambridge Planetary Science, pp. 68-101). Cambridge: Cambridge University Press.